\documentclass[english]{article}
\usepackage[T1]{fontenc}
\usepackage[latin9]{inputenc}
\usepackage{wrapfig}
\usepackage{amsmath}
\usepackage{amsthm}
\usepackage{amssymb}

\makeatletter
\theoremstyle{plain}
\newtheorem{thm}{\protect\theoremname}
\theoremstyle{plain}
\newtheorem{lem}[thm]{\protect\lemmaname}
\theoremstyle{plain}
\newtheorem{cor}[thm]{\protect\corollaryname}
\theoremstyle{definition}
\newtheorem{defn}[thm]{\protect\definitionname}
\theoremstyle{plain}
\newtheorem{fact}[thm]{\protect\factname}
\theoremstyle{plain}
\newtheorem{question}[thm]{\protect\questionname}

\usepackage{todonotes}
\usepackage{microtype}
\usepackage{fullpage}
\usepackage{wrapfig}

\@ifundefined{showcaptionsetup}{}{%
 \PassOptionsToPackage{caption=false}{subfig}}
\usepackage{subfig}
\makeatother

\usepackage{babel}
\providecommand{\corollaryname}{Corollary}
\providecommand{\definitionname}{Definition}
\providecommand{\factname}{Fact}
\providecommand{\lemmaname}{Lemma}
\providecommand{\questionname}{Question}
\providecommand{\theoremname}{Theorem}

\begin{document}
\global\long\def\sharpETH{\mathsf{\#ETH}}%
\global\long\def\sharpWone{\mathsf{\#W[1]}}%
\global\long\def\sharpP{\mathsf{\#P}}%
\global\long\def\PerfMatch{\#\mathrm{PerfMatch}}%
\global\long\def\hadw{\eta}%

\title{Parameterizing the Permanent:\\
Hardness for $K_{8}$-minor-free graphs}
\author{Radu Curticapean\thanks{IT University of Copenhagen and Basic Algorithms Research Copenhagen (BARC), Copenhagen, Denmark. BARC is supported by Villum Foundation Grant No.~16582.}$\qquad$Mingji
Xia\thanks{State Key Laboratory of Computer Science, Institute of Software, Chinese Academy of Sciences,
University of Chinese Academy of Sciences, Beijing, China.
Supported by NSFC~61932002.}\date{}}
\maketitle
\begin{abstract}
In the 1960s, statistical physicists discovered a fascinating algorithm
for counting perfect matchings in \emph{planar} graphs. Valiant later
showed that the same problem is $\sharpP$-hard for \emph{general}
graphs. Since then, the algorithm for planar graphs was extended to
bounded-genus graphs, to graphs excluding $K_{3,3}$ or $K_{5}$,
and more generally, to any graph class excluding a fixed minor $H$
that can be drawn in the plane with a single crossing. This stirred
up hopes that counting perfect matchings might be polynomial-time
solvable for graph classes excluding \emph{any} fixed minor $H$.
Alas, in this paper, we show $\sharpP$-hardness for $K_{8}$-minor-free
graphs by a simple and self-contained argument.
\end{abstract}

\section{Introduction}

A \emph{perfect matching} in a graph $G$ is an edge-subset $M\subseteq E(G)$
such that every vertex of $G$ has exactly one incident edge in $M$.
Counting perfect matchings is a central and very well-studied problem
in counting complexity. It already starred in Valiant's seminal paper~\cite{Valiant1979a}
that introduced the complexity class $\sharpP$, where it was shown
that counting perfect matchings is $\sharpP$-complete. The problem
has driven progress in approximate counting and underlies the so-called
holographic algo\-rithms~\cite{Valiant2008,Cai2010a,Cai2007,Cai2008}.
It also occurs outside of counting complexity, e.g., in statistical
physics, via the partition function of the \emph{dimer model}~\cite{Temperley1961,Kasteleyn1961,Kasteleyn1967}.
In algebraic complexity theory, the matrix \emph{permanent} is a very
well-studied algebraic variant of the problem of counting perfect
matchings~\cite{Agrawal2006}. Evaluating permanents is equivalent
to counting perfect matchings in bipartite graphs: Given a bipartite
input graph $G$ on $n+n$ vertices with its $n\times n$ bi-adjacency
matrix $A$, the permanent $\mathrm{per}(A)$ counts exactly the perfect
matchings in $G$.

\subsection*{Algorithms for restricted graph classes}

A long line of research, dating back to the 1960s, identified structural
restrictions on $G$ that facilitate the problem of counting perfect
matchings.  Statistical physics gave the first examples of useful
restrictions: On the graphs $G$ of \emph{regular lattices}, it turns
out that perfect matchings are counted by the determinants of highly
structured matrices whose eigenvalues can be derived explicitly~\cite{Temperley1961,Kasteleyn1961}.
This approach was later generalized to arbitrary \emph{planar graphs}
$G$: More precisely, and very surprisingly, it is possible to flip
the signs of some entries in the adjacency matrix of a planar graph
$G$ to obtain a matrix $A$ such that $\sqrt{\det(A)}$ counts the
perfect matchings in $G$~\cite{Kasteleyn1967}. The entries to be
flipped are determined by a so-called \emph{Pfaffian orientation}
of $G$, which can be computed in linear time for planar graphs. Overall,
a polynomial-time algorithm for counting perfect matchings in planar
graphs follows, the so-called \emph{FKT method }(an acronym for the
names of their inventors Fisher, Kasteleyn, and Temperley).

Little~\cite{Little1974} and Vazirani~\cite{Vazirani1989} later
generalized the FKT method from planar graphs to the more general
class of graphs excluding $K_{3,3}$ as a minor. Such graphs can be
obtained inductively by ``gluing together'' planar graphs and $K_{5}$.
Little showed that $K_{3,3}$-free graphs still admit a Pfaffian orientation
by combining Pfaffian orientations of the individual parts, and Vazirani
later obtained a polynomial-time and poly-logarithmic space algorithm
for finding such an orientation.

Still working with Pfaffian orientations, it was shown by Gallucio
and Loebl~\cite{Galluccio1998} and Tesler~\cite{Tesler2000} that
perfect matchings can be counted in time $4^{g}n^{O(1)}$ for graphs
$G$ that are embedded on a surface of genus $g$. In other words,
the problem is fixed-parameter tractable in the parameter $g$. These
algorithms use Pfaffian orientations to express the number of perfect
matchings in $G$ as a linear combination of $4^{g}$ determinants.
A simplified algorithm by the authors~\cite{Curticapean} bypasses
the explicit use of Pfaffian orientations and instead reduces in a
black-box manner to $4^{g}$ instances of counting perfect matchings
in planar graphs.

The mold of Pfaffian orientations was broken by Straub, Thierauf and
Wagner~\cite{Straub2014}, and Curticapean~\cite{Curticapean2014},
who independently designed polynomial-time algorithms for counting
perfect matchings in graphs excluding a $K_{5}$-minor. As such graphs
do not necessarily admit Pfaffian orientations, a different algorithmic
approach was needed: In hindsight, the new algorithms transferred
the protrusion replacement technique from parameterized complexity~\cite{DBLP:conf/focs/BodlaenderFLPST09}
into the counting setting.

Moreover, a standard dynamic programming approach shows that counting
perfect matchings is fixed-parameter tractable in graphs of bounded
tree-width $t$. This can be improved to $2^{t}n^{O(1)}$ time~\cite{DBLP:conf/esa/RooijBR09},
where the base $2$ is optimal under the strong exponential-time hypothesis~\cite{DBLP:conf/soda/CurticapeanM16}.

\subsection*{Towards excluding general fixed minors}

We observe that any tractable graph class listed above excludes some
fixed minor $H$. That is, starting from a graph $G$ in the class,
it is not possible to obtain $H$ by deleting edges/vertices and contracting
edges. For example, planar graphs exclude $K_{3,3}$ and $K_{5}$,
bounded-genus graphs exclude sufficiently large complete graphs, and
bounded-treewidth graphs even exclude large grids. It is therefore
natural to ask whether we can count perfect matchings in \emph{any}
graph class excluding fixed minors. 

On the positive side, Curticapean~\cite{Curticapean2014} and Eppstein
and Vazirani~\cite{DBLP:conf/spaa/EppsteinV19} lifted the algorithms
for $K_{3.3}$-minor-free and $K_{5}$-minor-free graphs to $H$-minor-free
graphs for any graph $H$ that can be drawn in the plane with a single
crossing, for example, $H=K_{3,3}$ and $H=K_{5}$. (Note that excluding
single-crossing minors $H$ yields different graph classes than the
class of single-crossing graphs themselves, for which an immediate
reduction to the FKT method is possible.) In fact, these algorithms
run in fixed-parameter tractable time $f(k)n^{O(1)}$ for some function
$f$ depending only on the size $k$ of an excluded single-crossing
minor. Note that the exponent of $n$ does not grow with $k$.

On the negative side, parameterized complexity rules out such fixed-parameter
tractable algorithms for counting perfect matchings in \emph{$k$-apex
graphs}~\cite{Curticapean}; these are the graphs that are planar
up to deleting $k$ vertices. More precisely, it was previously shown
by the authors that counting perfect matchings is $\sharpWone$-hard
on $k$-apex graphs, suggesting that algorithms for this problem require
$n^{g(k)}$ time for $g\in\omega(1)$. As $k$-apex graphs exclude
$K_{k+5}$-minors, algorithms for counting perfect matchings in $K_{t}$-free
graphs in turn require $n^{g(t)}$ time for $g\in\omega(1)$. It could
however still be possible that, for any fixed $t\in\mathbb{N}$, some
polynomial-time algorithm counts perfect matchings in $K_{t}$-free
graphs, with an exponent depending on $t$.

In fact, such algorithms seemed within close reach: Robertson and
Seymour's graph structure theorem~\cite{DBLP:journals/jct/RobertsonS03a}
shows that graphs $G$ excluding a fixed $H$-minor can be obtained
as clique-sums of particular graphs that are near-embeddable on surfaces
of fixed genus. Very roughly speaking, this means that $G$ is glued
together from certain pieces, similar to graphs excluding $K_{3,3}$
or $K_{5}$ but with some ``upgrades''. These upgrades involve raising
the \emph{genus} in the decomposition pieces (from $0$ to some constant
depending on $H$), adding a constant number of \emph{apex} vertices
(vertices that can connect arbitrarily to the remainder of the piece),
and adding a constant number of \emph{vortices} (graphs of bounded
path-width that are aligned with the boundary of a face).

\begin{wrapfigure}{o}{0.5\columnwidth}%
\centering
\begingroup%
  \makeatletter%
  \providecommand\color[2][]{%
    \errmessage{(Inkscape) Color is used for the text in Inkscape, but the package 'color.sty' is not loaded}%
    \renewcommand\color[2][]{}%
  }%
  \providecommand\transparent[1]{%
    \errmessage{(Inkscape) Transparency is used (non-zero) for the text in Inkscape, but the package 'transparent.sty' is not loaded}%
    \renewcommand\transparent[1]{}%
  }%
  \providecommand\rotatebox[2]{#2}%
  \newcommand*\fsize{\dimexpr\f@size pt\relax}%
  \newcommand*\lineheight[1]{\fontsize{\fsize}{#1\fsize}\selectfont}%
  \ifx\svgwidth\undefined%
    \setlength{\unitlength}{200.80964096bp}%
    \ifx\svgscale\undefined%
      \relax%
    \else%
      \setlength{\unitlength}{\unitlength * \real{\svgscale}}%
    \fi%
  \else%
    \setlength{\unitlength}{\svgwidth}%
  \fi%
  \global\let\svgwidth\undefined%
  \global\let\svgscale\undefined%
  \makeatother%
  \begin{picture}(1,0.60007668)%
    \lineheight{1}%
    \setlength\tabcolsep{0pt}%
    \put(0,0){\includegraphics[width=\unitlength,page=1]{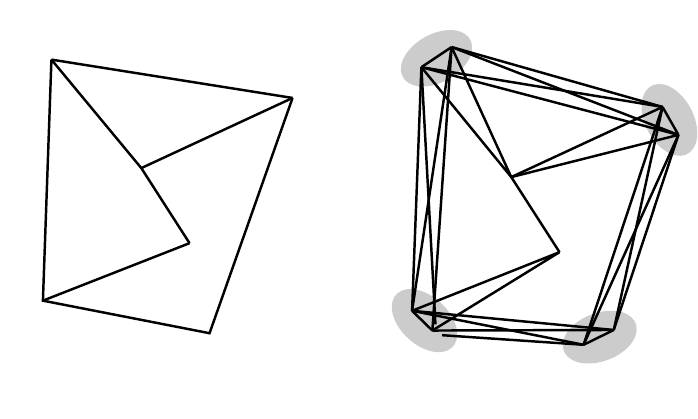}}%
    \put(0.05291427,0.54631808){\makebox(0,0)[lt]{\lineheight{1.25}\smash{\begin{tabular}[t]{l}$u$\end{tabular}}}}%
    \put(0.51564486,0.49473751){\makebox(0,0)[lt]{\lineheight{1.25}\smash{\begin{tabular}[t]{l}$u^1$\end{tabular}}}}%
    \put(0.58529855,0.56048017){\makebox(0,0)[lt]{\lineheight{1.25}\smash{\begin{tabular}[t]{l}$u^2$\end{tabular}}}}%
    \put(0,0){\includegraphics[width=\unitlength,page=2]{ringblowup.pdf}}%
    \put(0.03497006,0.10244814){\makebox(0,0)[lt]{\lineheight{1.25}\smash{\begin{tabular}[t]{l}$v$\end{tabular}}}}%
    \put(0.52552222,0.0662635){\makebox(0,0)[lt]{\lineheight{1.25}\smash{\begin{tabular}[t]{l}$v^2$\end{tabular}}}}%
    \put(0.48920566,0.13200625){\makebox(0,0)[lt]{\lineheight{1.25}\smash{\begin{tabular}[t]{l}$v^1$\end{tabular}}}}%
  \end{picture}%
\endgroup%

\caption{\label{fig: ringblowups}A ring blowup.}
\end{wrapfigure}

Counting perfect matchings supports most of these upgrades: The problem
is fixed-parameter tractable in bounded-genus graphs, while apex vertices
can be handled by brute-force in $n^{O(1)}$ time, and the gluing
operation can be handled as in the simpler case of excluded single-crossing
minors. Therefore, it only remains to handle vortices. However, even
a minimal example of vortices was unresolved: We say that a \emph{ring
blowup} is a graph obtained from a drawing of a planar graph by cloning
each vertex on the outer face into two copies, as shown in the left
part of Figure~\ref{fig: ringblowups}. In the terminology of the
graph structure theorem, ring blowups are certain planar graphs with
a single vortex. Progress towards polynomial-time algorithms for counting
perfect matchings in $H$-minor-free graphs was halted because such
algorithms were not even known for ring blowups.

\subsection*{Our results}

We show that counting perfect matchings remains $\sharpP$-hard in
ring blowups. Known general results in graph minor theory~\cite{JORET201361}
then already imply the existence of a graph $H$ such that counting
perfect matchings is $\sharpP$-hard in $H$-minor-free graphs. The
particularly simple structure of ring blowups however allows us to
exclude $K_{8}$-minors from them in a self-contained way. (Note that
ring blowups \emph{can} contain $K_{7}$-minors, as Figure~\ref{fig: simplify-minors}
shows.) We then obtain our main theorem:
\begin{thm}
\label{thm: main-thm}Counting perfect matchings is $\sharpP$-hard
in graphs excluding $K_{8}$-minors.
\end{thm}

A weaker version of this theorem was previously announced in a survey
on parameterized counting~\cite{DBLP:conf/iwpec/Curticapean18}.
To prove Theorem~\ref{thm: main-thm}, we reduce from the $\sharpP$-hard
problem of counting perfect matchings in a graph $G$. Our reduction
hinges upon a particular \emph{sign-crossing gadget} (see Figure~\ref{fig: sign-crossing})
that can be used to remove crossings at the cost of disrupting the
perfect matching count: After inserting a sign-crossing gadget between
crossing edges $e,f\in E(G)$, perfect matchings that contain \emph{both}
$e$ and $f$ are counted with a factor $-1$, and only perfect matchings
containing \emph{at most one} of $e$ or $f$ are counted properly.
Such sign-crossing gadgets were previously used in the theory of matchgates~\cite{Cai2014}
and in the hardness proof for counting perfect matchings in $k$-apex
graphs~\cite{Curticapean}. In our proof, sign-crossings are used
to transform $G$ into a ring blowup while preserving the perfect
matching count. By a surprisingly simple construction, we can ensure
that sign-crossings come in equivalent pairs, so that any $-1$ factors
introduced by sign-crossings cancel via $(-1)^{2}=1$.

\section{Preliminaries\label{sec:Preliminaries}}

To give a self-contained proof of Theorem~\ref{thm: main-thm}, we
first state some preliminaries from counting complexity and graph
minor theory. Graphs will be undirected and may be edge-weighted;
we implicitly consider $w:E(G)\to\mathbb{Q}$ to be the weight function.
Given a vertex $v\in V(G)$, we write $I(v)$ for the set of edges
incident with $v$. Furthermore, given a set $S\subseteq V(G)$, we
write $G[S]$ for the subgraph of $G$ induced by $S$. 

\subsection{Counting perfect matchings and gadgets}

We define counting problems as functions $\mathrm{\#A}:\{0,1\}^{*}\to\mathbb{Q}$,
where inputs (graphs, formulas, numbers) are implicitly encoded as
bitstrings. For example, $\mathrm{\#SAT}$ asks to count the satisfying
assignments to Boolean formulas. Likewise, when given as input a graph
with edge-weights from a constant-sized\footnote{The assumption of $|W|=\mathcal{O}(1)$ ensures that $\PerfMatch(G)$
can be represented with polynomially many bits.} set $W\subseteq\mathbb{Q}$, the problem $\PerfMatch$ asks to determine
the quantity
\begin{equation}
\PerfMatch(G)=\sum_{\substack{M\subseteq E(G)\,\mathrm{is}\,\mathrm{a}\\
\mathrm{perfect}\,\mathrm{matching}
}
}\prod_{e\in M}w(e).\label{eq: perfmatch}
\end{equation}

We say that $\mathrm{\#A}$ admits a polynomial-time Turing reduction
to $\mathrm{\#B}$ if $\mathrm{\#A}$ can be solved in polynomial
time with an oracle for $\mathrm{\#B}$, and we say that $\mathrm{\#B}$
is $\sharpP$-hard if $\mathrm{\#SAT}$ admits such a reduction to
$\mathrm{\#B}$. Our hardness proofs are based on the following theorem:
\begin{figure}
\subfloat[\label{fig: insert-crossing}The sign-crossing gadget is inserted
into a crossing between edges $e,f$. The gadget is drawn into a small
disk to avoid introducing further crossings. We call $e_{1},e_{2},f_{1},f_{2}$
the \emph{external edges} of the gadget.]{\centering
\begingroup%
  \makeatletter%
  \providecommand\color[2][]{%
    \errmessage{(Inkscape) Color is used for the text in Inkscape, but the package 'color.sty' is not loaded}%
    \renewcommand\color[2][]{}%
  }%
  \providecommand\transparent[1]{%
    \errmessage{(Inkscape) Transparency is used (non-zero) for the text in Inkscape, but the package 'transparent.sty' is not loaded}%
    \renewcommand\transparent[1]{}%
  }%
  \providecommand\rotatebox[2]{#2}%
  \newcommand*\fsize{\dimexpr\f@size pt\relax}%
  \newcommand*\lineheight[1]{\fontsize{\fsize}{#1\fsize}\selectfont}%
  \ifx\svgwidth\undefined%
    \setlength{\unitlength}{188.80011749bp}%
    \ifx\svgscale\undefined%
      \relax%
    \else%
      \setlength{\unitlength}{\unitlength * \real{\svgscale}}%
    \fi%
  \else%
    \setlength{\unitlength}{\svgwidth}%
  \fi%
  \global\let\svgwidth\undefined%
  \global\let\svgscale\undefined%
  \makeatother%
  \begin{picture}(1,0.53690038)%
    \lineheight{1}%
    \setlength\tabcolsep{0pt}%
    \put(0,0){\includegraphics[width=\unitlength,page=1]{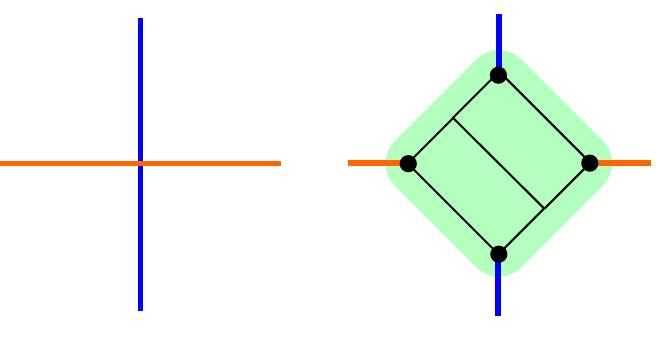}}%
    \put(0.24496535,0.45792428){\color[rgb]{0,0,1}\makebox(0,0)[lt]{\lineheight{1.25}\smash{\begin{tabular}[t]{l}$e$\end{tabular}}}}%
    \put(0.02966651,0.3241819){\color[rgb]{1,0.4,0}\makebox(0,0)[lt]{\lineheight{1.25}\smash{\begin{tabular}[t]{l}$f$\end{tabular}}}}%
    \put(0.78036211,0.49089425){\color[rgb]{0,0,1}\makebox(0,0)[lt]{\lineheight{1.25}\smash{\begin{tabular}[t]{l}$e_1$\end{tabular}}}}%
    \put(0.78422871,0.05936497){\color[rgb]{0,0,1}\makebox(0,0)[lt]{\lineheight{1.25}\smash{\begin{tabular}[t]{l}$e_2$\end{tabular}}}}%
    \put(0.51505943,0.2146494){\color[rgb]{1,0.4,0}\makebox(0,0)[lt]{\lineheight{1.25}\smash{\begin{tabular}[t]{l}$f_1$\end{tabular}}}}%
    \put(0.96665858,0.21464942){\color[rgb]{1,0.4,0}\makebox(0,0)[lt]{\lineheight{1.25}\smash{\begin{tabular}[t]{l}$f_2$\end{tabular}}}}%
    \put(0.74851186,0.32814863){\color[rgb]{0,0,0}\rotatebox{-43.194688}{\makebox(0,0)[lt]{\lineheight{1.25}\smash{\begin{tabular}[t]{l}$-1$\end{tabular}}}}}%
    \put(0,0){\includegraphics[width=\unitlength,page=2]{sign-crossing.pdf}}%
  \end{picture}%
\endgroup%
}$\qquad\quad\ $\subfloat[\label{fig: crossing-states}The top row shows, up to symmetry in
the last two cases, the possible configurations with $0,2$ or $4$
external edges. The middle row shows the possible extensions by matchings
within the gadget. The bottom row lists the total weighted sums of
these extensions.]{\centering
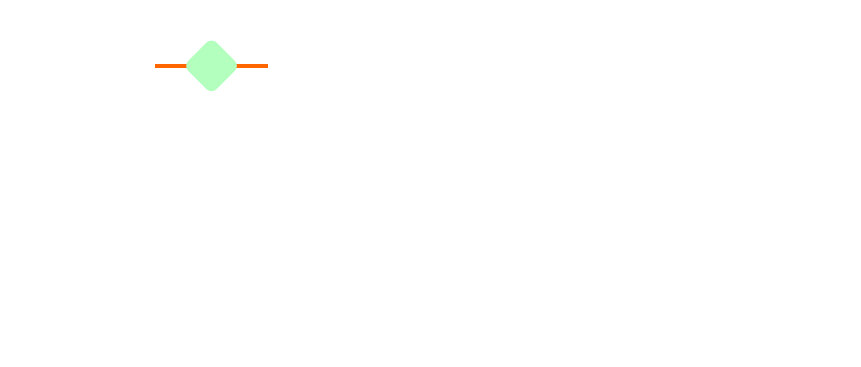}

\caption{\label{fig: sign-crossing}The planar sign-crossing gadget introduced
by Cai and Gorenstein~\cite{Cai2014}.}
\end{figure}

\begin{thm}[\cite{Valiant1979a,Dagum1992}]
\label{thm:Counting-perfect-matchings}Counting perfect matchings
is $\sharpP$-hard, even for unweighted $3$-regular graphs.
\end{thm}

The problem of counting perfect matchings behaves well under insertion
of gadgets; in the context of this problem, gadgets are called \emph{matchgates}.
For example, given a drawing of a not necessarily planar graph $G$
with a crossing involving edges $e,f\in E(G)$, we can replace the
crossing by the planar \emph{sign-crossing gadget} from Cai and Gorenstein~\cite{Cai2014},
as shown in Figure~\ref{fig: insert-crossing}. The resulting graph
essentially counts perfect matchings in $G$, but with a significant
twist: Any perfect matching $M\subseteq E(G)$ containing both $e$
and $f$ is weighted by an additional factor of $-1$. In other words,
every perfect matching $M$ incurs a factor 
\begin{align}
\chi_{e,f}(M) & :=\begin{cases}
-1 & \{e,f\}\subseteq M,\\
1 & \text{otherwise.}
\end{cases}\label{eq: chi_ef}
\end{align}

Note that a ``perfectly planarizing'' crossing gadget that does
not introduce negative signs would render counting perfect matchings
polynomial-time solvable by reduction to the FKT method. It can even
be shown unconditionally that no such gadget exists~\cite{Cai2014}.

The claimed functionality of the sign-crossing gadget follows from
standard techniques in the area of so-called Holant problems, see~\cite{Valiant2008,Cai2014,Curticapean2015}.
In the following, we give a self-contained proof.
\begin{lem}
\label{lem: sign-crossing}Let $G$ be a weighted graph that is drawn
in the plane with crossing edges $e,f\in E(G)$ of weight $1$, and
let $G'$ be obtained by inserting a sign-crossing gadget as in Figure~\ref{fig: insert-crossing}.
Then, with $\chi_{e,f}$ as in (\ref{eq: chi_ef}), we have
\[
\PerfMatch(G')=\sum_{\substack{M\subseteq E(G)\,\mathrm{is}\,\mathrm{a}\\
\mathrm{perfect}\,\mathrm{matching}
}
}\chi_{e,f}(M)\prod_{e\in M}w(e).
\]
\end{lem}

\begin{proof}
Let $X=\{e_{1},e_{2},f_{1},f_{2}\}$ be the external edges of the
sign-crossing gadget $S$ inserted at $e,f$. Let $G''=G-V(S)$ be
the graph obtained from $G$ by removing all gadget vertices. For
a subset $T\subseteq X$ of external edges, write $V(T)$ for the
endpoints of edges in $T$. We observe that any perfect matching $M\subseteq E(G')$
can be obtained by first choosing a subset $T\subseteq X$ and afterwards
extending $T$ by \emph{independently} choosing perfect matchings
in $S-V(T)$ and $G''-V(T)$. It follows that
\begin{equation}
\PerfMatch(G')=\sum_{T\subseteq X}\left(\prod_{e\in T}w(e)\right)\cdot\underbrace{\PerfMatch(S-V(T))}_{=:f(T)}\,\cdot\ \PerfMatch(G''-V(T)).\label{eq: perfmatch-types}
\end{equation}

We observe that any set $T\subseteq X$ with $f(T)\neq0$ has even
cardinality, as $S-V(T)$ would otherwise have an odd number of vertices,
and hence, no perfect matchings. The values of $f(T)$ for sets $T\subseteq X$
of even cardinality are calculated in Figure~\ref{fig: crossing-states}:
Each column shows such a set $T$ in the top row and lists the perfect
matchings of $S-V(T)$ below. The value $\PerfMatch(S-V(T))$ is then
obtained in the bottom row as the $\pm1$-weighted count of the listed
perfect matchings.

The calculations from Figure~\ref{fig: crossing-states} show that
any set $T\subseteq X$ with $f(T)\neq0$ is \emph{consistent} in
the sense that it includes none/both of $\{e_{1},e_{2}\}$ and none/both
of $\{f_{1},f_{2}\}$. Given such a consistent set $T$, define $\tilde{T}\subseteq\{e,f\}$
by forgetting the subscripts in $T$: Include $e$ into $\tilde{T}$
iff $\{e_{1},e_{2}\}\subseteq T$, and likewise for $f$. The term
in (\ref{eq: perfmatch-types}) corresponding to $T$ counts precisely
those perfect matchings $M$ in $G$ with $M\cap\{e,f\}=\tilde{T}$,
except that $f(T)$ introduces a factor of $-1$ if $\tilde{T}=\{e,f\}$.
This proves the lemma.
\end{proof}
If several crossings are replaced by sign-crossing gadgets, an inductive
application of Lemma~\ref{lem: sign-crossing} shows that the factors
introduced by sign-crossing gadgets multiply. A particularly interesting
situation occurs when edges are drawn as curves rather than straight
lines, as two edges $e$ and $f$ may then cross more than once. Our
corollary supports this case, which will prove very useful in the
next section.
\begin{cor}
\label{cor: multiple-crossings}Let $G$ be an unweighted graph that
is drawn in the plane. Choose $t\in\mathbb{N}$ crossings and write
$e_{i},f_{i}\in E(G)$ for the edges involved in the $i$-th crossing.
Let $G'$ be obtained by inserting a sign-crossing gadget at each
of the $t$ chosen crossings. Then we have
\[
\PerfMatch(G')=\sum_{\substack{M\subseteq E(G)\,\mathrm{is}\,\mathrm{a}\\
\mathrm{perfect}\,\mathrm{matching}
}
}\prod_{i=1}^{t}\chi_{e_{i},f_{i}}(M).
\]
\end{cor}

\subsection{Graph minor theory}

A graph $H$ is a \emph{minor} of $G$, written $H\preceq G$, if
$H$ can be obtained by repeated edge deletions and contractions and
vertex deletions. This is equivalent to the existence of a minor model
of $H$ in $G$:
\begin{defn}
\label{def:minor-model}A \emph{minor model} of $H$ in $G$ is a
collection of pairwise disjoint branch sets $S_{v}\subseteq V(G)$
for $v\in V(H)$ such that (i) each set $S_{v}$ for $v\in V(H)$
induces a connected subgraph of $G$, and (ii) for every edge $uv\in E(H)$,
some edge of $G$ runs between $S_{u}$ and $S_{v}$.
\end{defn}

For example, the colored sets in Figure~\ref{fig: simplify-minors}
show minor models of $K_{7}$ in ring blowups. The \emph{Hadwiger
number} $\eta(G)$ of a graph $G$ is the maximum $k\in\mathbb{N}$
with $K_{k}\preceq G$.

A plane graph is a planar graph that is given together with a concrete
planar embedding. We define a particular graph class from plane graphs
by ``blowing up'' their outer faces, see Figure~\ref{fig: ringblowups}.

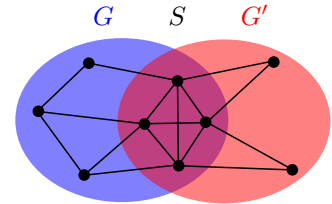
\begin{wrapfigure}{o}{0.3\columnwidth}%
\centering
\begingroup%
  \makeatletter%
  \providecommand\color[2][]{%
    \errmessage{(Inkscape) Color is used for the text in Inkscape, but the package 'color.sty' is not loaded}%
    \renewcommand\color[2][]{}%
  }%
  \providecommand\transparent[1]{%
    \errmessage{(Inkscape) Transparency is used (non-zero) for the text in Inkscape, but the package 'transparent.sty' is not loaded}%
    \renewcommand\transparent[1]{}%
  }%
  \providecommand\rotatebox[2]{#2}%
  \newcommand*\fsize{\dimexpr\f@size pt\relax}%
  \newcommand*\lineheight[1]{\fontsize{\fsize}{#1\fsize}\selectfont}%
  \ifx\svgwidth\undefined%
    \setlength{\unitlength}{120.82823181bp}%
    \ifx\svgscale\undefined%
      \relax%
    \else%
      \setlength{\unitlength}{\unitlength * \real{\svgscale}}%
    \fi%
  \else%
    \setlength{\unitlength}{\svgwidth}%
  \fi%
  \global\let\svgwidth\undefined%
  \global\let\svgscale\undefined%
  \makeatother%
  \begin{picture}(1,0.61054081)%
    \lineheight{1}%
    \setlength\tabcolsep{0pt}%
    \put(0,0){\includegraphics[width=\unitlength,page=1]{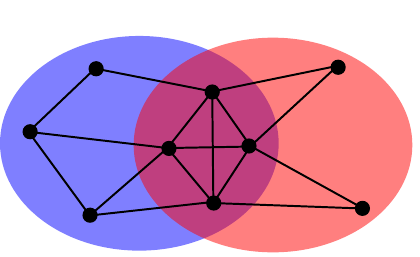}}%
    \put(0.24285713,0.56729365){\color[rgb]{0,0,1}\makebox(0,0)[lt]{\lineheight{1.25}\smash{\begin{tabular}[t]{l}$G$\end{tabular}}}}%
    \put(0.47756752,0.56729365){\makebox(0,0)[lt]{\lineheight{1.25}\smash{\begin{tabular}[t]{l}$S$\end{tabular}}}}%
    \put(0.70303243,0.56729365){\color[rgb]{1,0,0}\makebox(0,0)[lt]{\lineheight{1.25}\smash{\begin{tabular}[t]{l}$G'$\end{tabular}}}}%
  \end{picture}%
\endgroup%

\caption{A clique-sum.}
\end{wrapfigure}%

\begin{defn}
\label{def: planar-blowup}Given a plane graph $\hat{Q}$ with outer
face $O$, the \emph{blowup} of $\hat{Q}$ is the graph obtained by
successively replacing each vertex $v\in O$ by two clones $v^{1},v^{2}$
having the same neighborhood as $v$, and then adding the edge $v^{1}v^{2}$.
We call $v^{1}$ and $v^{2}$ \emph{blowup vertices}. A \emph{ring
blowup} is any subgraph of the blowup $Q$ of a plane graph $\hat{Q}$.
We also call $\hat{Q}$ a \emph{reduct} of $Q$.
\end{defn}

The notion of clique-sums will feature prominently in Section~\ref{sec: adorned}.
\begin{defn}
Let $G$ and $G'$ be two graphs that share a common vertex-set $S=V(G)\cap V(G')$,
which is a clique in $G$ and $G'$. A \emph{clique-sum} $G\oplus_{S}G'$
is any graph that can be obtained from the union $G\cup G'$ by deleting
some edges with both endpoints in $S$.
\end{defn}

A standard separator argument bounds the Hadwiger number of clique-sums
by that of their constituents, see \cite[Lemma~2.1]{JORET201361}
for a proof.
\begin{fact}
\label{fact: clique-sum}For any graphs $G,G'$ and any clique-sum
$G\oplus_{S}G'$ for $S=V(G)\cap V(G')$, we have $\hadw(G\oplus_{S}G')\leq\max\{\hadw(G),\hadw(G')\}$.
\end{fact}

For a proof sketch, note that no fixed minor model of $K_{t}$ can
place some branch sets entirely within $V(G)\setminus S$ and others
entirely within $V(G')\setminus S$. It follows that all branch sets
intersect $V(G)$ or all intersect $V(G')$. In the first case, we
can delete all vertices from $V(G')\setminus S$ without losing edges
between branch sets, since $S$ is a clique. The second case is symmetric.

\section{Reducing to ring blowups}

In this section, we show how to transform any unweighted graph $G$
into a ring blowup while preserving the value of $\PerfMatch$. The
main idea, spelled out in Lemma~\ref{lem: ringblow} and illustrated
in Figure~\ref{fig: transform}, is to arrange the vertices of $G$
on a circle and then bend the edges of $G$ to push crossings across
the perimeter of the circle, where blowups will take care of them.
As an edge $e$ is bent, it will introduce new crossings with other
edges $g$, but our construction ensures that any edge $g$ crossed
while bending $e$ is crossed \emph{exactly twice}. When we then introduce
sign-crossing gadgets at these crossings, any $-1$ factors from gadgets
are guaranteed to come in pairs, so the overall product of these factors
is $1$. Hence, going from $G$ to $G'$, the value of $\PerfMatch$
is preserved via Corollary~\ref{cor: multiple-crossings}. In Lemma~\ref{lem: remove weights},
we then use a standard reduction in counting complexity to remove
the $-1$ weights introduced into $G'$ by sign-crossings.
\begin{lem}
\label{lem: ringblow}Let $G$ be an unweighted graph with $n$ vertices
and $m$ edges. In polynomial time, we can construct a ring blowup
$G'$ on $O(n+m^{3})$ vertices and edge-weights $\pm1$ such that
$\PerfMatch(G)=\PerfMatch(G')$ holds and all edges incident with
blowup vertices of $G'$ have weight $1$.
\end{lem}

\begin{proof}
As shown in Figure~\ref{fig: trans1}, we first place $V(G)$ on
a circle $C$ in the plane and draw the edges of $G$ as straight
lines inside of $C$. The placement is chosen such that no three edges
intersect in the same point and every ray from the center to the perimeter
of $C$ contains at most one point that is a crossing or vertex of
$G$. Both conditions can be ensured by perturbing an arbitrary placement
of $V(G)$ on $C$.

The circle divides the plane into two regions; we call the induced
subgraphs of $G$ contained in these regions (both including $C$)
the \emph{outer }and\emph{ inner} part of $G$. Initially, the outer
part only contains $C$. 

Let $s\in O(m^{2})$ be the number of crossings in our drawing of
$G$ and let $P_{1},\ldots,P_{s}\in\mathbb{R}^{2}$ be their locations.
For each $i\in[s]$, shoot a ray from the center of $C$ to $P_{i}$
and write $\ell_{i}$ for the segment of this ray from $P_{i}$ to
the perimeter of $C$. Note that distinct rays are disjoint and contain
no vertices of $G$. For $i\in[s]$ in sequence, write $e_{i},f_{i}\in E(G)$
for the edges involved in crossing $P_{i}$, write $m_{i}$ for the
number of edges crossed by segment $\ell_{i}$ and enumerate the crossed
edges as $g_{i,1},\ldots,g_{i,m_{i}}\in E(G)$. We bend $e_{i}$
and $f_{i}$ in a sufficiently narrow neighborhood of $\ell_{i}$
to cross $C$, as shown in the middle part of Figure~\ref{fig: trans2}.
For any $j\in[m_{i}]$, this process adds two crossings between $e_{i},g_{i,j}$
and two crossings between $f_{i},g_{i,j}$.
\begin{figure}
\subfloat[\label{fig: trans1}Initial drawing of $G$ with circle $C$ (dashed)
that induces an inner part (shaded) and outer part. The segments $\ell_{i}$
extending from crossings to $C$ are drawn as red lines.]{\centering
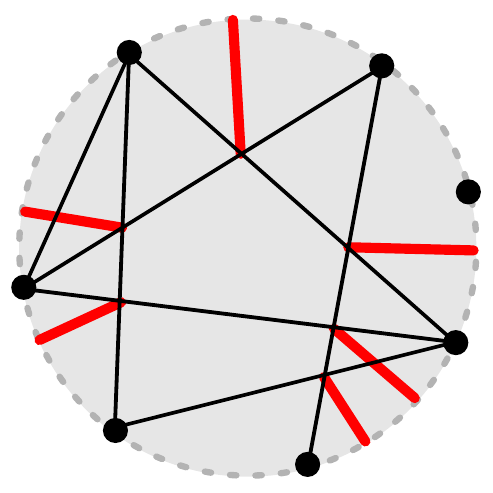}$\qquad$\subfloat[\label{fig: trans2}Left: The $i$-th crossing, involving edges $e_{i},f_{i}$,
and the edges $g_{i,j}$ crossed by segment $\ell_{i}$. Middle: The
edges $e_{i}$ and $f_{i}$ are bent towards the outer part along
$\ell_{i}$. Each edge $g_{i,j}$ now crosses each of $e_{i}$ and
$f_{i}$ twice. Right: Inserting sign-crossing gadgets and dragging
the top vertices $v_{1},\ldots,v_{4}$ of outermost crossing gadgets
onto $C$. The dark ellipses show that $G'$ is a ring blowup.]{\centering
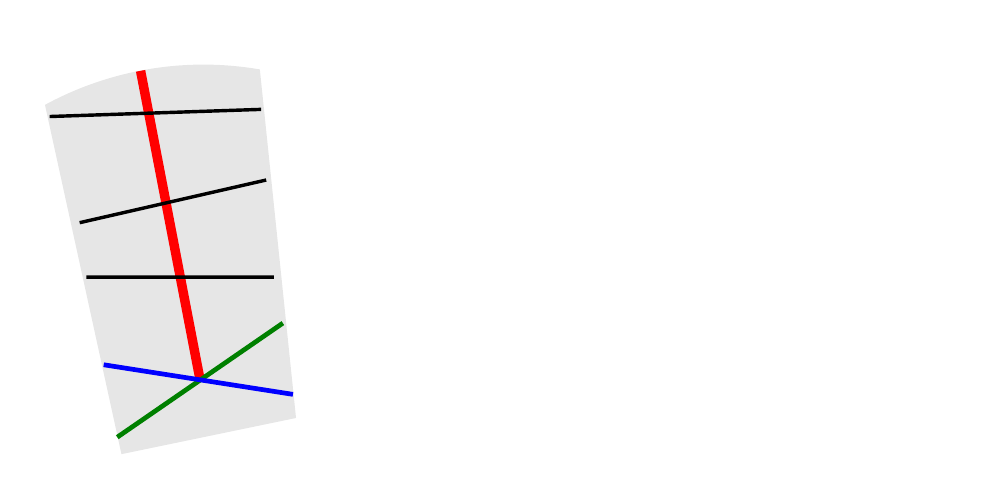}\caption{\label{fig: transform}How Lemma~\ref{lem: ringblow} moves crossings
to the outer part of $G$ to obtain a ring blowup $G'$.}
\end{figure}

After all original crossings $P_{1},\ldots,P_{s}$ are processed,
we insert a sign-crossing gadget at each crossing in the inner part
of $G$, as shown in the right part of Figure~\ref{fig: trans2}.
Note that no sign-crossing gadgets are inserted at crossings in the
outer part. To simplify the subsequent argument, we drag some vertices
of the sign-crossing gadgets onto $C$, as shown in Figure~\ref{fig: trans2}.
Overall, we obtain a new graph $G'$ with edge-weights $1$ and $-1$,
and we obtain from Corollary~\ref{cor: multiple-crossings} that
\begin{align}
\PerfMatch(G') & =\sum_{M}\prod_{i=1}^{s}\prod_{j=1}^{m_{i}}\underbrace{\chi_{e_{i},g_{i,j}}^{2}(M)}_{=1}\cdot\underbrace{\chi_{f_{i},g_{i,j}}^{2}(M)}_{=1}=\PerfMatch(G).\label{eq: perfmatch-crossdrag}
\end{align}

Taking inventory, while going from $G$ to $G'$, we replaced all
crossings from the inner part with planar gadgets and added no new
crossings to the inner part, as different segments $\ell_{i},\ell_{j}$
are disjoint. Via sign-crossings, we added $O(\sum_{i}m_{i})=O(sm)=O(m^{3})$
vertices into the inner part, where we recall that $m_{i}$ is the
number of crossings between segment $\ell_{i}$ and the edges in $G$.
Each crossing is contained in the outer part and involves edges $v_{1}v_{3}$
and $v_{2}v_{4}$ for some consecutive block of vertices $v_{1},\ldots,v_{4}$
on the circle $C$. The vertex blocks of different crossings are disjoint,
so we can define a reduct of $G'$ by identifying $v_{1}=v_{2}$ and
$v_{3}=v_{4}$ in each block, as shown in Figure~\ref{fig: trans2}.
Then also no edges of weight $-1$ are incident with blowup vertices.
This shows that $G'$ is a ring blowup satisfying the specifications
of the lemma.
\end{proof}
To conclude this section, we remark that negative edge-weights can
be removed from the graphs constructed before while staying in the
graph class of ring blowups.
\begin{lem}
\label{lem: remove weights}The problem $\PerfMatch$ restricted to
the graphs from Lemma~\ref{lem: ringblow} (that is, ring blowups
with edge-weights $\pm1$ such that edges of weight $-1$ are not
incident with blowup vertices) admits a polynomial-time Turing reduction
to $\PerfMatch$ in unweighted ring blowups.
\end{lem}

A standard proof of this lemma, see e.g.~\cite{Curticapean2015},
replaces occurrences of the weight $-1$ with an indeterminate~$x$;
this turns the number of perfect matchings in an $n$-vertex graph
into a polynomial $p\in\mathbb{Z}[x]$ of degree at most $d=n/2$.
This polynomial can be evaluated at non-negative integer inputs $0,\ldots,d$
via planar gadgets, and the value $p(-1)$ can then be recovered from
$p(0),\ldots,p(d)$ via polynomial interpolation. As Lemma~\ref{lem: ringblow}
guarantees that edges of weight $-1$ are not incident with blowup
vertices, the planar gadgets introduced in Lemma~\ref{lem: remove weights}
can be contained within the inner part of the resulting graphs.

Combining Theorem~\ref{thm:Counting-perfect-matchings} (the $\sharpP$-hardness
of counting perfect matchings in $3$-regular graphs) with Lemmas~\ref{lem: ringblow}~and~\ref{lem: remove weights},
we immediately obtain:
\begin{thm}
\label{thm: hard-blowups}The problem $\PerfMatch$ is $\sharpP$-hard
in unweighted ring blowups.
\end{thm}

In fact, our proof even shows that $\PerfMatch$ is $\sharpP$-hard
in graphs $G$ that are obtained from plane graphs by adding edges
$v_{1}v_{3}$ and $v_{2}v_{4}$ between disjoint blocks of consecutive
vertices $v_{1},\ldots,v_{4}$ on the outer face. For the arguments
in Section~\ref{sec: adorned}, we will however need fully general
ring blowups.

\section{Bounding the Hadwiger number\label{sec: adorned}}

In this section, we bound the Hadwiger number of ring blowups by $7$.
First, we show in Lemma~\ref{lem: make-simple} that it suffices
to consider \emph{simple ring blowups}, that is, blowups of plane
graphs that have all but $\leq3$ vertices on the outer face. In Lemma~\ref{lem: reduce-complication},
we iteratively remove certain \emph{complications} from simple ring
blowups. When this process terminates, we obtain graphs that can be
handled by trivial arguments in Lemma~\ref{lem: complication-free}.
\begin{defn}
A \emph{simple ring }is a plane graph $Q$ with outer face $O$ such
that $Q-O$ is a complete graph with $\leq3$ vertices. We call $W=V(Q)\setminus O$
the \emph{inner face} of $Q$.\footnote{Technically, this need not be a face.}
A \emph{simple ring blowup} is any subgraph of the blowup of a simple
ring. (Equivalently, a graph is a simple ring blowup if it has a simple
ring as reduct.)
\end{defn}

\begin{figure}
$\qquad$\subfloat[\label{fig: simplify-minors-1}The initial graph $G$ with a minor
model of $K_{7}$, indicated by vertex colors. For clarity, edges
contained in branch sets are also colored.]{\centering
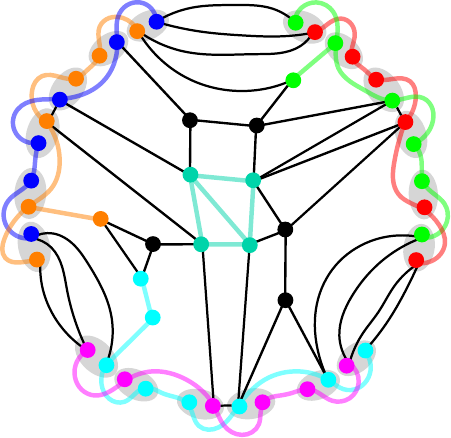}$\qquad$\subfloat[\label{fig: simplify-minors-2}Steps~1-2 removed vertices not contained
in branch sets and contracted the turquoise branch set, which contains
no blowup vertices.]{\centering
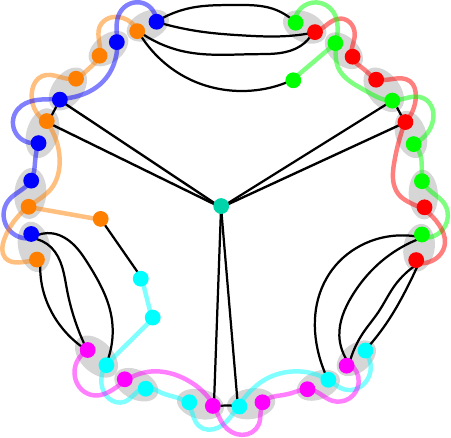}$\qquad$\subfloat[\label{fig: simplify-minors-3}Step~3 successively contracted all
edges that run between blowup vertices and non-blowup vertices. The
result is a simple ring blowup.]{\centering
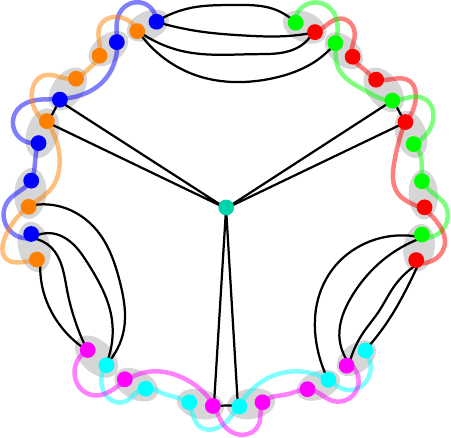}

\caption{\label{fig: simplify-minors}How Lemma~\ref{lem: make-simple} cleans
up a ring blowup while maintaining a $K_{t}$-minor.}
\end{figure}

In the following lemma, we adapt an argument by Joret and Wood~\cite[Lemma 3.4]{JORET201361}
to make ring blowups simple without decreasing their Hadwiger number.
See Figure~\ref{fig: simplify-minors} for an illustration of the
process; the last drawing shows an example of a simple ring blowup
with a single vertex on the inner face. The lemma repeatedly uses
the fact that, given a minor model of a graph $H$ in another graph
$G$, contracting an edge contained a branch set canonically induces
a minor model of $H$ in the resulting graph. 

\begin{lem}
\label{lem: make-simple}For any ring blowup $G$, there is a simple
ring blowup $Q$ with $\hadw(G)\leq\eta(Q)$.
\end{lem}

\begin{proof}
We abbreviate $t=\hadw(G)$. The lemma holds for $t\leq4$, since
there clearly are simple ring blowups containing $K_{4}$-minors.
We may therefore assume $t\geq5$ in the following. 

Let $\hat{G}$ be a reduct of $G$, with outer face $O$, and let
$G$ be drawn as a blowup of $\hat{G}$. Fix a minor model $S_{1},\ldots,S_{t}\subseteq V(G)$
of $K_{t}$ in $G$, define $G':=G$ and proceed as follows.
\begin{enumerate}
\item Delete from $G'$ all vertices not contained in $S_{1}\cup\ldots\cup S_{t}$.
Then the sets $S_{1},\ldots,S_{t}$ still yield a minor model of $K_{t}$
in the resulting graph, still called $G'$, as no edges between branch
sets were deleted.
\item Contract every branch set $S_{i}$ that does not contain blowup vertices.
This yields a minor model of $K_{t}$ in the resulting graph, still
called $G'$. The $\ell\leq t$ vertices $W=\{w_{1},\ldots,w_{\ell}\}$
resulting from the contraction induce a planar drawing of $K_{\ell}$,
so we have $\ell\leq4$. We may even assume $\ell=3$: Otherwise,
if $\ell=4$, then one of the vertices, say $w_{4}$, would be enclosed
by the cycle on $W\setminus\{w_{4}\}$ in the drawing of $G'$. But
then $w_{4}$ cannot have edges to the $t-\ell$ other branch sets,
so the $t$ overall branch sets cannot form a minor model of $K_{t}$
for $t\geq5$.
\item For every edge $uv$ fully contained in a branch set, where $u$ is
a blowup vertex and $v$ is not, contract $uv$ into $u$. This still
yields a minor model of $K_{t}$ in the resulting graph $G'$.
\end{enumerate}
Summing up, we see that $G'$ contains a $K_{t}$-minor, so it suffices
to prove that $G'$ is a simple ring blowup. Note that the $\leq3$
vertices in $W$ form a clique in $G'$, and all other vertices are
blowup vertices. By applying the operations used to transform $G$
into $G'$ on the reduct of $G$, we obtain a reduct of $G'$ that
is a simple ring, thus proving the lemma.
\end{proof}
By removing certain structures that we call ``complications'', simple
ring blowups can be simplified even further. In the following, it
will be useful to work with the reducts.
\begin{figure}
\centering
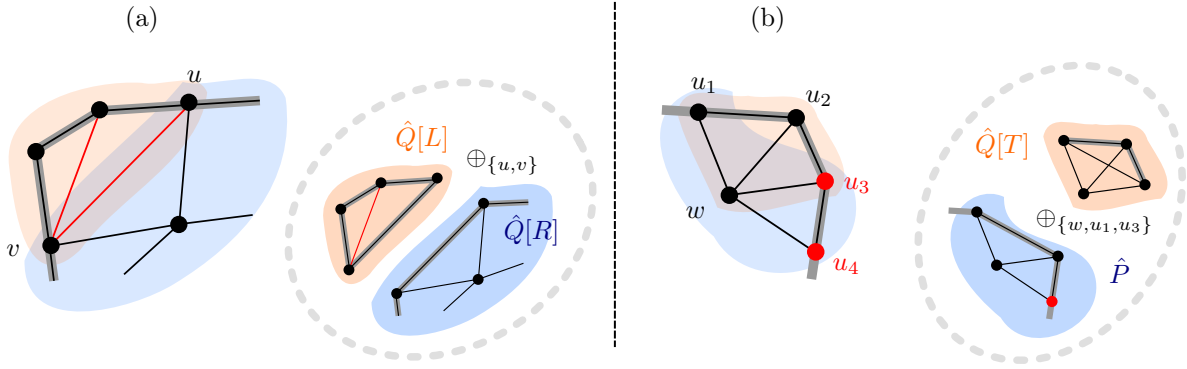

\caption{\label{fig: complications}This figure exemplifies complications in
a simple ring $\hat{Q}$ and illustrates the main steps in the proof
of Lemma~\ref{lem: reduce-complication}. The outer face of $\hat{Q}$
is drawn as a thick line. When ignoring shading and the encircled
drawings, the figure illustrates complications of both types (marked
red). The shading and encircled drawings indicate the clique-sums
arising in the proof of Lemma~\ref{lem: reduce-complication}.}
\end{figure}

\begin{defn}
Given a simple ring $\hat{Q}$ with outer face $O$ and inner face~$W$,
a \emph{complication} in $\hat{Q}$ is 
\begin{enumerate}
\item[(a) ] any edge between vertices $u,v\in O$ that are not consecutive in
the cyclic order of $O$, or
\item[(b) ] given a vertex $w\in W$, any neighbor $o\in O$ of $w$ after its
first two neighbors in the cyclic order of $O$.
\end{enumerate}
\end{defn}

Figure~\ref{fig: complications} illustrates complications of the
different types. We show in the following lemma that it suffices to
bound the Hadwiger number of the blowups of triangulated complication-free
simple rings. Here, we say that a graph is \emph{triangulated} if
every face but its outer face is a triangle.
\begin{lem}
\label{lem: reduce-complication}If $\hadw(Z)\leq7$ holds for the
blowup $Z$ of any triangulated and complication-free simple ring
$\hat{Z}$, then $\hadw(Q)\leq7$ holds for any simple ring blowup
$Q$.
\end{lem}

\begin{proof}
Let $\hat{Q}$ be a simple ring with complications, outer face $O$,
and inner face $W$. Let $Q$ be the blowup of $\hat{Q}$. We may
assume $\hat{Q}$ to be triangulated; this can be ensured by adding
edges, which does not decrease $\eta(Q)$. 

The goal is to decompose $\hat{Q}$ along clique-sums into simple
rings with strictly less complications; the triangulation will be
kept intact along the way. Using Fact~\ref{fact: clique-sum} inductively,
it will therefore suffice to bound $\eta(Z)$ when $Z$ is the blowup
of a triangulated and complication-free ring $\hat{Z}$. The following
notation will be useful: For vertex sets $S\subseteq V(\hat{Q})$,
let $B(S)$ be obtained by replacing each vertex $v\in S\cap O$ with
$v^{1},v^{2}$.

We start by removing all type-a complications. To this end, consider
a type-a complication involving $u,v\in O$. Then there exist sets
$L,R\subseteq V(\hat{Q})$ such that $\hat{Q}$ is a clique-sum $\hat{Q}[L]\oplus_{S}\hat{Q}[R]$,
as shown in the left part of Figure~\ref{fig: complications}. This
in turn means that the blowup $Q$ is a clique-sum $Q[B(L)]\oplus_{B(S)}Q[B(R)]$.
Via Fact~\ref{fact: clique-sum}, it suffices to show that $Q[B(L)]$
and $Q[B(R)]$ are $K_{8}$-minor-free. But $Q[B(L)]$ and $Q[B(R)]$
are the blowups of $\hat{Q}[L]$ and $\hat{Q}[R]$, which are triangulated
simple rings with at least one type-a complication less. By induction,
we may therefore assume in the following that all type-a complications
are processed.

Now consider a type-b complication, with $w\in W$ and consecutive
neighbors $u_{1},u_{2},u_{3}\in O$ of $w$ in the cyclic order of
$O$. Note that we may indeed assume the neighbors to be consecutive,
since all type-a complications were processed. Let us define $S=\{w,u_{1},u_{3}\}$
and $T=S\cup\{u_{2}\}$. Then $\hat{Q}$ is a clique-sum $\hat{P}\oplus_{S}\hat{Q}[T]$
for the simple ring $\hat{P}$ obtained from $\hat{Q}$ by removing
$u_{2}$ and adding the edge $u_{1}u_{3}$, as shown in the right
part of Figure~\ref{fig: complications}. This in turn implies that
$Q$ is a clique-sum $P\oplus_{B(S)}Q[B(T)]$, where $P$ is the blowup
of $\hat{P}$. We observe that $Q[B(T)]$ has only $7$ vertices,
so it suffices to exclude $K_{8}$ from $P$, where $\hat{P}$ is
a triangulated simple ring that still has no type-a complications,
and one type-b complication less. By induction, we may therefore assume
that $\hat{Q}$ has no complications at all. This proves the lemma.
\end{proof}
Finally, an elementary case distinction allows us to handle complication-free
ring blowups.
\begin{lem}
\label{lem: complication-free}For any triangulated and complication-free
simple ring $\hat{Z}$ with blowup $Z$, we have $\eta(Z)\leq7$.
\end{lem}

\begin{proof}
Let $O$ and $W$ with $|W|\leq3$ be the outer and inner faces of
$\hat{Z}$. Every edge $e\in E(\hat{Z})$ either has both endpoints
in $W$, one endpoint in $O$ and $W$ each, or both endpoints of
$e$ lie consecutively on $O$ (since $Z$ has no type-a complications.)
If $W$ is empty, then we have $|V(\hat{Z})|\leq3$, since $Z$ is
triangulated. This already implies $\hadw(Z)\leq|V(Z)|\leq6$.

If $W$ is non-empty, then any vertex $w\in W$ has $\leq2$ neighbors
in $O$, since $\hat{Z}$ has no type-b complications. Furthermore,
since $\hat{Z}$ is triangulated, any pair of vertices in $W$ shares
a neighbor in $O$, because $\hat{Z}$ would otherwise contain a chordless
cycle of length $4$. This implies the following:
\begin{itemize}
\item If $|W|\leq2$, then $|O|\leq2$ and thus $\hadw(Z)\leq|V(Z)|\leq6$. 
\item If $|W|=3$, then $\hat{Z}$ and its blowup $Z$ are the following
graphs, where $|V(Z)|=9$ and $|E(Z)|=30$.\\
\begingroup%
  \makeatletter%
  \providecommand\color[2][]{%
    \errmessage{(Inkscape) Color is used for the text in Inkscape, but the package 'color.sty' is not loaded}%
    \renewcommand\color[2][]{}%
  }%
  \providecommand\transparent[1]{%
    \errmessage{(Inkscape) Transparency is used (non-zero) for the text in Inkscape, but the package 'transparent.sty' is not loaded}%
    \renewcommand\transparent[1]{}%
  }%
  \providecommand\rotatebox[2]{#2}%
  \newcommand*\fsize{\dimexpr\f@size pt\relax}%
  \newcommand*\lineheight[1]{\fontsize{\fsize}{#1\fsize}\selectfont}%
  \ifx\svgwidth\undefined%
    \setlength{\unitlength}{425.19685039bp}%
    \ifx\svgscale\undefined%
      \relax%
    \else%
      \setlength{\unitlength}{\unitlength * \real{\svgscale}}%
    \fi%
  \else%
    \setlength{\unitlength}{\svgwidth}%
  \fi%
  \global\let\svgwidth\undefined%
  \global\let\svgscale\undefined%
  \makeatother%
  \begin{picture}(1,0.23272535)%
    \lineheight{1}%
    \setlength\tabcolsep{0pt}%
    \put(0,0){\includegraphics[width=\unitlength,page=1]{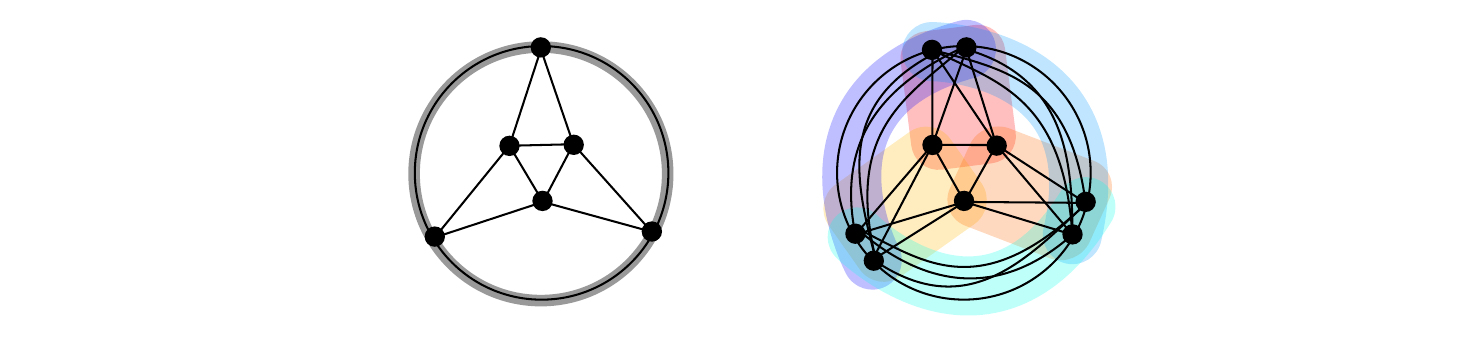}}%
    \put(0.23115725,0.11010985){\makebox(0,0)[lt]{\lineheight{1.25}\smash{\begin{tabular}[t]{l}$\hat Z$\end{tabular}}}}%
    \put(0.51525287,0.11010985){\makebox(0,0)[lt]{\lineheight{1.25}\smash{\begin{tabular}[t]{l}$Z$\end{tabular}}}}%
  \end{picture}%
\endgroup%
\\
To obtain a $K_{8}$-minor from $Z$, one vertex must be deleted or
one edge must be contracted. Deleting a vertex removes at least $6$
edges. Contracting an edge reduces the number of edges by least $3$,
since every edge in $Z$ is contained in some $K_{4}$-subgraph (shown
above as colored blobs) and contracting an edge of $K_{4}$ yields
a $K_{3}$, thus losing $3$ edges. It follows that any $8$-vertex
minor of $Z$ has at most $27<|E(K_{8})|$ edges and therefore cannot
be $K_{8}$.
\end{itemize}
This covers all cases for $|W|$, thus proving the lemma.
\end{proof}
The proof of Theorem~\ref{thm: main-thm} is now immediate.
\begin{proof}[Proof of Theorem~\ref{thm: main-thm}]
By Theorem~\ref{thm: hard-blowups}, the problem $\PerfMatch$ is
$\sharpP$-hard in unweighted ring blowups $G$. It remains to show
that $G$ excludes $K_{8}$-minors. By Lemma~\ref{lem: make-simple},
we have $\eta(G)\leq\eta(Q)$ for some simple ring blowup $Q$. By
Lemma~\ref{lem: reduce-complication}, we have $\eta(Q)\leq7$ if
$\eta(Z)\leq7$ holds for all blowups of triangulated and complication-free
ring graphs $\hat{Z}$, which in turn is true by Lemma~\ref{lem: complication-free}.
\end{proof}

\section{Conclusion and outlook\label{sec:Conclusion-and-outlook}}

We showed that the FKT method for planar graphs cannot be extended
to graphs excluding arbitrary fixed minors. However, our work leaves
open an exhaustive classification of the minors whose exclusion renders
$\PerfMatch$ polynomial-time solvable. This is not an artifact of
our analysis: As Figure~\ref{fig: simplify-minors} shows, the graphs
constructed by our reduction \emph{can} contain $K_{7}$-minors, so
our reduction inherently fails to address the open case of $K_{7}$-minor-free
$\PerfMatch$. This prompts the obvious question:
\begin{question}
What is the complexity of $\PerfMatch$ in graphs excluding $K_{6}$
or $K_{7}$? More generally, given any fixed graph $H$, what is the
complexity of $H$-minor-free $\PerfMatch$?
\end{question}

Turning towards a bigger picture, it is also interesting to investigate
which other counting problems benefit from excluded minors. This can
be studied systematically in the framework of \emph{Holant problems},
of which counting perfect matchings constitutes a representative example.

In a future version of this paper, we rule out $\exp(o(\sqrt{n}))$
time algorithms for $\PerfMatch$ with edge-weights $\pm1$ under
the exponential-time hypothesis. This lower bound introduces various
complications that we eschewed here in favor of a self-contained and
simple presentation. Note that graphs excluding fixed minors have
tree-width $O(\sqrt{n})$, and therefore standard algorithms for counting
perfect matchings in graphs of bounded tree-width yield matching $\exp(O(\sqrt{n}))$
time upper bounds on $H$-minor-free graphs.

\bibliographystyle{plain}
\bibliography{C:/Users/Radu/OneDrive/Research/MainBib}

\end{document}